# Intrinsic nonlinear Hall effect and gate-switchable Berry curvature sliding in twisted bilayer graphene

Meizhen Huang[1,#], Zefei Wu[1,5,#,*], Xu Zhang[2,#], Xuemeng Feng[1], Zishu Zhou[1], Shi Wang[1], Yong Chen[1,3], Chun Cheng[3], Kai Sun[4], Zi Yang Meng[2,*], Ning Wang[1,*]

[1]Department of Physics and Center for Quantum Materials, The Hong Kong University of Science and Technology, Hong Kong, China
[2]Department of Physics and HKU-UCAS Joint Institute of Theoretical and Computational Physics, The University of Hong Kong, Hong Kong, China
[3]Department of Materials Science and Engineering, Southern University of Science and Technology, Shenzhen, China
[4]Department of Physics, University of Michigan, Ann Arbor, Michigan 48109, USA
[5]Present address: Department of Physics and Astronomy and National Graphene Institute, University of Manchester, Manchester M13 9PL, UK.

[#]These authors contributed equally: Meizhen Huang, Zefei Wu, Xu Zhang.
*Corresponding author. Email: phwang@ust.hk; zymeng@hku.hk; zefei.wu@manchester.ac.uk

## Abstract

Though the observation of the quantum anomalous Hall effect and nonlocal transport response reveals nontrivial band topology governed by the Berry curvature in twisted bilayer graphene, some recent works reported nonlinear Hall signals in graphene superlattices that are caused by the extrinsic disorder scattering rather than the intrinsic Berry curvature dipole moment. In this work, we report a Berry curvature dipole induced intrinsic nonlinear Hall effect in high-quality twisted bilayer graphene devices. We also find that the application of the displacement field substantially changes the direction and amplitude of the nonlinear Hall voltages, as a result of a field-induced sliding of the Berry curvature hotspots. Our work not only proves that the Berry curvature dipole could play a dominant role in generating the intrinsic nonlinear Hall signal in graphene superlattices with low disorder densities, but also demonstrates twisted bilayer graphene to be a sensitive and fine-tunable platform for second harmonic generation and rectification.

## I. INTRODUCTION

In the past few years, moiré superlattices, formed by stacking two layers of two-dimensional materials with a small twist angle, have led to many intriguing discoveries[1,2]. For example, unconventional superconductivity and correlated insulating states have been observed in flatbands of twisted bilayer graphene (TBG)[3-5], which offers a promising new platform to study strongly correlated many-body phenomena. In addition to correlation effects, moiré bands are also believed to offer nontrivial band topology[6-8], which is supported by the observation of ferromagnetism and quantum anomalous Hall effect in TBG aligned with BN, where the single spin- and valley-resolved miniband has a nontrivial Chern number[9-11], and nonlocal resistance in superlattice gaps of TBG originated from two nontrivial bulk $Z_2$ invariants[12]. While previous studies mostly focus on the global (or topological) properties of electronic wave functions, our knowledge about the *k*-space distribution of the Berry curvature remains limited. Studying Berry curvature properties within flat bands is of particular interest because it plays an important role not only in generating topological transport, but also in stabilizing correlated insulating states[13] and determing superfluid stiffness[14].

The nonlinear Hall effect (NHE) is the generation of a second harmonic Hall voltage in response to the injection current under time-reversal symmetric conditions[15]. Previous observations of NHE in low symmetry crystals[16,17], strained two-dimensional materials[18,19], and twisted transition metal dichalcoginides[20] have attracted great interest because of the potential applications, including frequency doubling and rectification[21,22]. Theoretical studies have shown that the NHE mainly has two possible microscopic origins. First, it can arise from a nonuniform distribution of the Berry curvature, also known as the Berry curvature dipole[15,23,24], and therefore becomes a powerful method to detect Berry curvatures in nonmagnetic quantum materials. Second, it can also arise from skew scattering[25,26], where the inherent chirality of the electron wave function induces transverse asymmetric scattering, which applies to a wider class of materials even without a Berry curvature dipole. In analogy to the intrinsic spin Hall effect which is caused by spin-orbit coupling in the band structure and survives in the limit of zero disorder, and the extrinsic spin Hall effect which is caused by spin-orbit coupling between Bloch electrons and impurities, here we call the Berry curvature dipole induced NHE as the intrinsic NHE, and the disorder-induced NHE as the extrinsic NHE. In graphene superlattices, the extrinsic NHE has been observed in graphene-BN superlattices and TBG[27,28], and intrinsic NHE has been recently observed in twisted double-bilayer graphene (TDBG) in the vicinity of the band gaps[29]. The former two works failed to observe any Berry curvature dipole behavior and the following one mainly focused on the gap properties instead of the flatband properties as Berry curvatures in TDBG predominantly reside in the vicinity of the band edges. Until now, whether the Berry curvature or the scattering would play a dominant role in graphene superlattices was an open question. A systematic study of the Berry curvature dipole induced NHE inside the flatbands is still missing.

Here we studied the NHE inside the flatbands of TBG with a twist angle of 1.3°. By using this angle away from the magic angle, the strong correlation effect declines while band dispersion in TBG keeps relatively flat. The distribution of Berry curvature is diffused in the band and there will be significant signal from Berry curvature dipole detected within band, which distinguishes with the nonlinear Hall in TDBG, where the Berry curvatures are mostly accumulated nearly full fillings. Upon injecting an ac current, a second harmonic Hall voltage that scales quadratically with the injection current together with a vanishing second harmonic longitudinal response is observed, demonstrating a robust NHE originated from Berry curvature dipole. Even if the first harmonic transport properties of our TBG samples remain relatively unchanged upon the applied displacement field[2,30], we find that displacement field can slide the Berry curvature hotspots and lead to significant changes in the direction and amplitude of the nonlinear Hall signal. Our theoretical calculation fully explains the experimental observation of such electrically tunable NHE and reveals its origin from the sliding of the Berry curvature hotspots on the flatbands of the moiré superlattice. Our observation therefore establishes TBG as a fine-tunable platform for applications in second harmonic generation and rectification.

## II. EXPERIMENTAL RESULTS

TBG heterostructures, as schematically shown in Fig. 1a, are fabricated by the "tear-and-stack" method[31]. A dual gate geometry is used to independently control the total carrier density $n$ and the displacement field $D$ (see Supplementary Note. 1 for details[32]). The scheme of longitudinal and transverse measurement is illustrated in Fig. 1b. We first characterize the TBG device by measuring the longitudinal resistance $R_{xx}$ and the Hall carrier density $n_H$ as a function of the gate voltage (or, equivalently, the filling factor $\upsilon$ of the bands) at a temperature of $T$ = 1.5 K. The sharp peak of $R_{xx}$ near the charge neutral point (CNP) indicates the mobility ~110,000 cm$^2$V$^{-1}$s$^{-1}$. In TBG, single-particle gaps form at CNP and at a superlattice density of $n = \pm n_s = \pm 4/A$, where $A$ is the moiré unit cell area and the factor 4 accounts for the spin and valley degeneracies. These fillings with single-particle gaps ($\upsilon = 0$ and $\pm 4$) can be identified in Fig. 1c (vertical dashed lines) by peaks of $R_{xx}$ and the sharp sign change in $n_H$, indicating the switching between electron like to hole like pockets as the Fermi energy swap through a band gap. While additional $R_{xx}$ peaks can be observed near the half filling and the quarter filling, $n_H$ at these positions shows no anomalies and follows $n_H = n$, suggesting that the correlation effect is relatively small in this device and a Bistritzer-MacDonald type of the flatband model[2], which we adapt in the theoretical calculation of NHE below, offers a good description of such a system. This behavior is also consistent with the observation of additional sign changes in $n_H$ at $3 < |\upsilon| < 4$ that matches the electron-hole charge carrier switching point (or van Hove singularities) in the single-particle band structure[39,40]. From the superlattice density $n_s = 4 \times 10^{12}$ cm$^{-2}$, the twist angle of this device is determined to be 1.3°. It is worthwhile to highlight that by utilizing this twisting angle away from the magic angle, complicated many-body effects, such as the anomalous enhancement of the nonlinear Hall signal from divergent effective mass[20], are avoided so that a clean platform where semiclassical description for NHE from band anomalous velocity can be used is obtained to study the Berry curvature dipole of intrinsic band structures. The advantage of such a flatband is that the distribution of Berry curvature is diffused in band so that there will be significant signal from Berry curvature dipole detected within band.

It is important to note that an ideal TBG is not expected to exhibit NHE due to the threefold rotational symmetry, which forces the Berry curvature dipole ($\Lambda$) to be zero[15]. However, the strain that arises in the sample fabrication process can break this symmetry and induce a nonzero nonlinear Hall signal[41,42]. As shown in the bottom panel of Fig. 1c, where the second harmonic Hall voltage $V_{\perp}^{2\omega}$ at $I^{\omega} = 100$ nA versus $\upsilon$ is plotted, nonzero responses are indeed observed. First of all, in agreement with the sign changes of $\Lambda$ at band gaps[42,43], $V_{\perp}^{2\omega}$ changes signs at CNP and full fillings. In addition, $V_{\perp}^{2\omega}$ show two peaks with opposite signs near the band gaps, which can be understood by the Berry curvature hotspots near the edges of the gap. To verify that $V_{\perp}^{2\omega}$ is indeed from Berry curvature contributions instead of extrinsic effects such as defect scattering, additional experiment data is provided in Supplementary Note. 2. The filling, displacement field, and temperature dependent data can be captured by the calculated Berry curvature dipole, but cannot be explained by the scattering-induced nonlinear Hall response.

We then systematically study the dependence of NHE with an out-of-plane electric displacement field *D* which induces charge and voltage differences between the layers (*D* field dependence of the first harmonic transport data shown in Supplementary Note. 3 indicate that the effect of displacement field on first harmonic transport of TBG is very weak and disorder's contribution to transport is very small.). Fig. 2a shows the map of $V_\perp^{2\omega}$ in the space of filling factor-displacement field (υ-*D*). $V_\perp^{2\omega}$ at υ = −1.5 and at υ = +1 are marked by the dashed line labeled with p and n in Fig. 2a, respectively. The magnitude and direction of $V_\perp^{2\omega}$ changes when we tune *D* while keeping υ unchanged. To be more specific, $V_\perp^{2\omega}$ at υ = −1.5 at various *D* is plotted in Fig. 2b. A region with small negative values of $V_\perp^{2\omega}$ is observed between two positive $V_\perp^{2\omega}$ peaks. Moreover, $V_\perp^{2\omega}$ shows a clear quadratic current–voltage (*I–V*) characteristic for all *D* (Fig. 2c), which, combined with the 2ω frequency, indicates a robust nonlinear Hall effect. Similar *D* switchable $V_\perp^{2\omega}$ is observed in the conduction band (Fig. 2d-e). In addition, *D* switchable $V_\perp^{2\omega}$ is observed in other devices with different twist angles, as shown in Supplementary Fig. 1. These observations show that *D* can effectively modulate the low-energy bands as well as the sign and the amplitude of the Berry curvature dipole *Λ*.

## III. BERRY CURVATURE SLIDING INDUCED SWITCHABLE DIPOLE

Numerical studies are performed to study the origin of the observed *D* field effect in the observed NHE results in TBG. Band structure of TBG under a uniaxial strain is calculated by using the Bistritzer-MacDonald model with a strain strength of 0.3% (the electric field is introduced by considering an electrostatic energy difference between two layers). Berry curvature *Ω* is calculated using the numerical method proposed in Ref. [44]. The *Λ* can be determined using $\Lambda = \int_k \left(\frac{\partial f_0}{\partial E}\right)\frac{\partial E}{\partial k}\Omega$, where $\int_k = d^2k/(2\pi)^2$, and $f_0$ is the fermion occupation number. In Fig. 3a, the calculated *Λ* as functions of *D* and υ is plotted (disorder induced extrinsic NHE is not included in this calculation and discussed separately in Supplementary Note 2 and 5). For a fixed υ, the calculated *Λ* shows an opposite sign for small *D* and large *D*, closely resembles the experimentally measured $V_\perp^{2\omega}$. In addition, experimental data shown in Fig. 1c can be captured by the theoretical calculation (See Supplementary Fig. 5). These consistencies verify that the observed electrically tunable NHE indeed results from *D* field tunable *Λ*.

To understand the microscopic origin, results of band structure and $\Omega$ of strained TBG at three different $D$ are shown in Fig. 3b. As shown by our calculation and the experimental observations, in the accessible range of $D$, single-particle gaps at CNP continuously present due to the breaking inversion symmetry in strained TBG. There is no band close-reopen happening which may change the sign of $\Omega$. Instead, when tuning $D$, the location of the $\Omega$ hotspot (indicated by the blue and red region) moves due to the change in the band structure. In Fig. 3c, the distribution of $\Omega \frac{\partial E}{\partial k}$ inside the reciprocal lattice at six representative points are plotted. The $\Lambda$ can then be determined using $\Lambda = \int_k \left(\frac{\partial f_0}{\partial E}\right) \frac{\partial E}{\partial k} \Omega$. For $\upsilon = +1$ at $D= +0.4$ (point n1) and $-0.4$V/nm (point n3), the dominated negative values of $\Omega \frac{\partial E}{\partial k}$ in the reciprocal lattice lead to a negative $\Lambda$. When $D$ is between these two values, at point n2, the location of the $\Omega$ hotspot slides into a band region that has a different sign of the band dispersion $\frac{\partial \varepsilon}{\partial k}$, leading to a sign change of $\Omega \frac{\partial E}{\partial k}$. Similarly, for the valence band (labeled with p1, p2 and p3), the predominantly values of $\Omega \frac{\partial E}{\partial k}$ at p1 and p3 is positive, and the sliding leads to negative values of $\Omega \frac{\partial E}{\partial k}$ at p2. Since $\Lambda = \int_k \left(\frac{\partial f_0}{\partial E}\right) \frac{\partial E}{\partial k} \Omega$, it is understandable that the sliding of $\Omega$ hotspot on band structure can induce sign changes of $\Lambda$.

## IV. DISCUSSION

To this end, we would like to comment on the two types of NHEs. It is proposed that both the Berry curvature dipole and the disorder scattering could induce a nonlinear Hall signal. The dipole contribution depends on the strain strength, and the disorder contribution depends on the disorder density. In TBG, according to theoretical calculation[45], an exponential decay form factor $\lambda(k, k + q + Q)$ with momentum transfer '$q + Q$' kills large momentum scattering tails out of several moiré Brillouin zones. Local disorder with carbon atom size has long momentum tail scaling as graphene Brillouin zone after Fourier transformation so that the form factor will sharply suppress the contribution from this kind of disorder (also the most common disorder). As a result, only spatially extended scatterer in moiré unit cell level (denoted as effective disorder) with small momentum scattering within the moiré Brillouin zone dominates in TBG's nonlinear transport. To compare the extrinsic (disorder) contribution with the intrinsic (dipole) contribution, Fig. 4 is provided, where screened charged Coulomb potential from effective disorder is used to compute the skew scattering contribution for NHE and detailed calculations can be found in Supplementary Note. 5. The nonlinear signal is dipole-dominated at a relatively large strain and a small effective disorder density (which is the case in our sample marked by star in Fig. 4 with a high mobility ~110,000 cm$^2$V$^{-1}$s$^{-1}$ and a effective disorder density ~2.7× 10$^6$ cm$^{-2}$), and scattering-dominated at the opposite (observed in TBG with mobility ~23,000 cm$^2$V$^{-1}$s$^{-1}$ and a disorder density ~3.4× 10$^8$ cm$^{-2}$[27]). Since the disorder strength can hardly be controlled, it is much easier and simpler to control the nonlinear Hall signal in high-quality TBG devices by manipulating the intrinsic dipole contribution via tuning the strain strength using flexible substrates[19].

In summary, we observed electrically tunable intrinsic NHE in TBG. In contrast to recent studies about the scattering-induced second harmonic generation in graphene superlattice[27,28], the NHE in our devices is induced by a nonzero Berry curvature dipole. In addition, our study indicates that the Berry curvature distribution in TBG is highly sensitive to the out-of-plane displacement field, which offers a highly efficient way to control or vary the amplitude and direction of the Berry curvature dipole, and thus the NHE. Our theoretical model calculation, where the sliding of the Berry curvature hotspots due to the interplay of strain, displacement field and flatband structure, semiquantitatively explains the experimental results. This work not only elucidates the intrinsic origin of the NHE in high-quality TBG to be the Berry curvature distribution inside the flatbands but also highlights TBG to be a fine-tunable platform for second harmonic generation and rectification.

## Acknowledgments

Grant support from the National Key R&D Program of China (2020YFA0309600) and the Research Grants Council (RGC) of Hong Kong (Project No. AoE/P701/20, 16303720, C6025-19G, and C7037-22G) are acknowledged. X. Z. and Z. Y. M. also acknowledge support from the RGC of Hong Kong (Project Nos. 17301420, 17301721, 17309822), the ANR/RGC Joint Research Scheme sponsored by RGC of Hong Kong and French National Research Agency (Project No. A_HKU703/22)), the K.C. Wong Education Foundation (Grant No. GJTD-2020-01). Device fabrication was performed at the MCPF and WMINST of HKUST with great technical support from Mr. LAI Chun Kit and Dr. Yuan Cai.

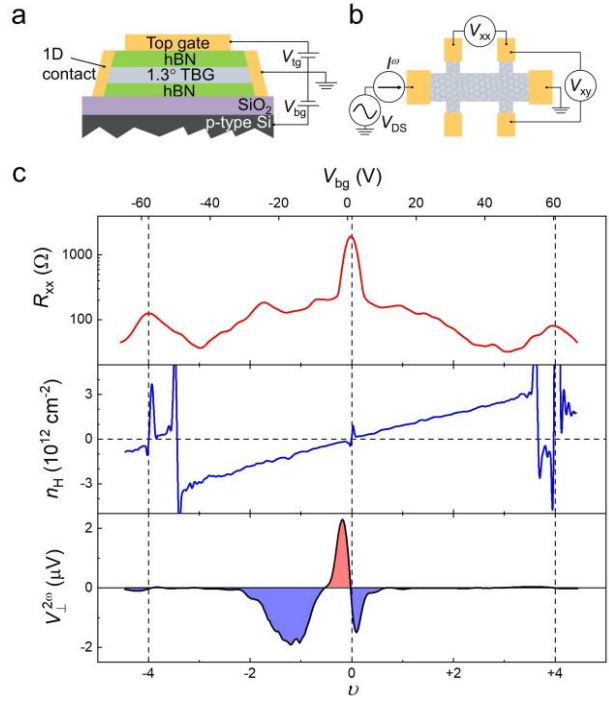

**Fig. 1 | Basic characterization of the TBG device. a,** Schematic structure of the 1.30° TBG device. The carrier density $n$ and the displacement field $D$ are controlled by the top and bottom gates with gate voltage $V_{tg}$ and $V_{bg}$, respectively. **b,** The four-probe measurement scheme. Here, we measure the longitudinal voltage $V_{xx}$ and Hall voltage $V_{xy}$ in the presence of an AC longitudinal current $I^\omega$. **c,** The longitudinal resistance $R_{xx} = V_{xx}/I^\omega$ (upper panel) and the carrier density measured via the Hall effect $n_H$ (middle panel) as functions of $\upsilon$ and $V_{bg}$ ($V_{tg}$=0 V). The vertical dashed lines denote the CNP and fully filled moiré bands. Bottom panel: The second harmonic Hall voltage $V_\perp^{2\omega}$ as a function of filling factor $\upsilon$ measured at $I^\omega = 100$ nA and $T = 1.5$ K.

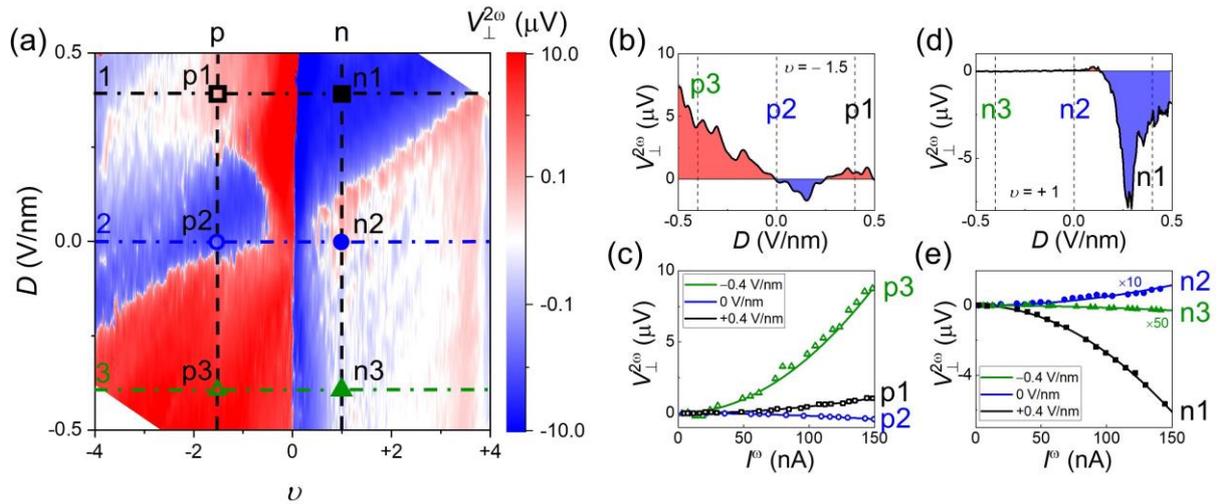

**Fig. 2 | Electric field tunable nonlinear Hall effect. a,** Filling and displacement field mapping of $V_\perp^{2\omega}$ measured at $T$ = 1.5 K. Dashed lines marked with p and n indicate $\upsilon = -1.5$ and $\upsilon = +1$, respectively. Dash dotted lines marked with 1, 2, and 3 indicate three different fields: $D = +0.4, 0, -0.4$ V/nm. Their intersection points are named as shown in the figure. **b,** $V_\perp^{2\omega}$ versus $D$ measured at $\upsilon = -1.5$ at $I^\omega = 100$ nA. **c,** $V_\perp^{2\omega}$ versus $I^\omega$ at various displacement fields at $\upsilon = -1.5$. Dots are the experimental data and the solid lines are parabolic fittings. **d, e,** Same as **b** and **c** for $\upsilon = +1$.

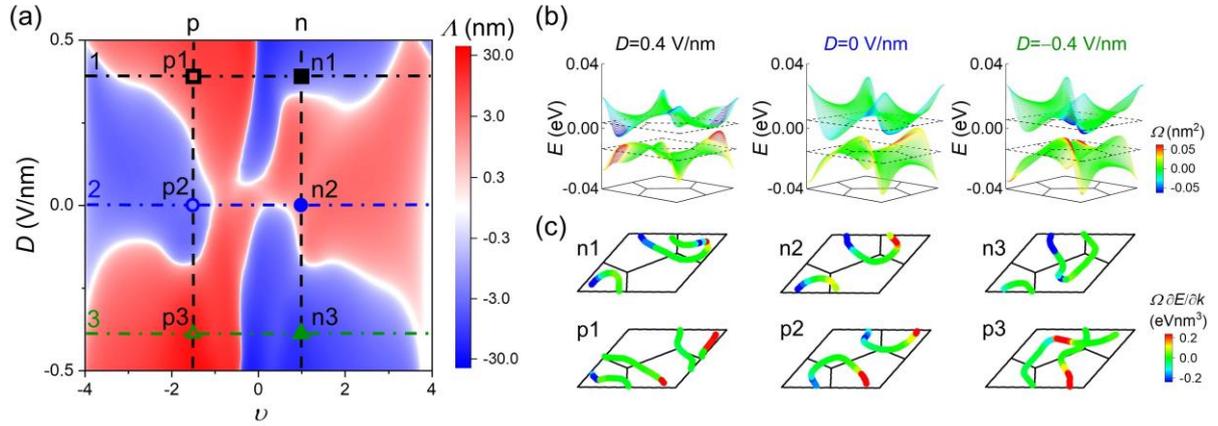

**Fig. 3 | Sliding of the Berry curvature hotspot. a,** $\Lambda$ as a function of $\upsilon$ and $D$ calculated by using the Bistritzer-MacDonald model with a strain strength of 0.3%. **b,** Calculated band structure of the moiré bands near charge neutrality with $D$= +0.4, 0, and -0.4V/nm. The color represents the value of $\Omega$. The parallelogram marks the reciprocal lattice (high symmetry points are labeled in Supplementary Fig. 9). **c,** The distribution of the product of the Berry curvature $\Omega$ and the slope of the band distribution $\frac{\partial E}{\partial k}$ within the reciprocal lattice at the energies marked by the dashed parallelograms in (**b**) at six representative points in (**a**). The zero-temperature dipole moment can be calculated by integrating $\Omega \frac{\partial E}{\partial k}$ at the Fermi level in the first moiré Brillouin zone.

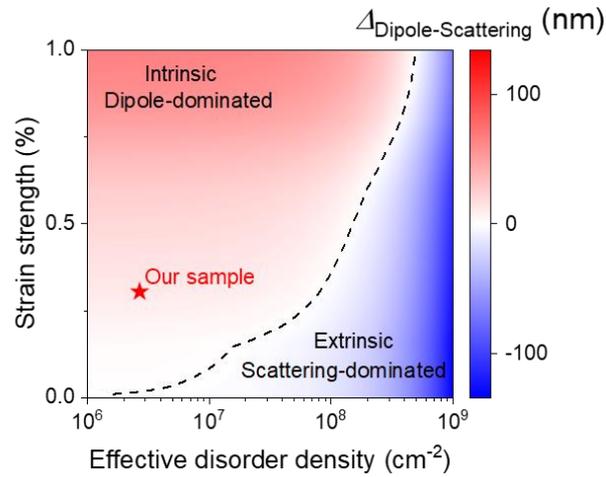

**Fig. 4 | Phase diagram of the NHE**. The color marks the difference between the Berry curvature dipole and the effective dipole of the scattering mechanism $\Delta_{\text{Dipole-scattering}}$. The dashed line represents the position where the Berry curvature dipole and the scattering contribute equally to the nonlinear Hall signal. The red (blue) area represents the dipole (scattering) dominated area. The strain strength is the total uniaxial strain used in theoretical calculation. The effective disorder is from screened Coulomb potential scatterer.

# Supplementary information

for

# Intrinsic nonlinear Hall effect and gate-switchable Berry curvature sliding in twisted bilayer graphene

Meizhen Huang[1,#], Zefei Wu[1,5,#,*], Xu Zhang[2,#], Xuemeng Feng[1], Zishu Zhou[1], Shi Wang[1], Yong Chen[1,3], Chun Cheng[3], Kai Sun[4], Zi Yang Meng[2,*], Ning Wang[1,*]

[1]Department of Physics and Center for Quantum Materials, The Hong Kong University of Science and Technology, Hong Kong, China
[2]Department of Physics and HKU-UCAS Joint Institute of Theoretical and Computational Physics, The University of Hong Kong, Hong Kong, China
[3]Department of Materials Science and Engineering, Southern University of Science and Technology, Shenzhen, China
[4]Department of Physics, University of Michigan, Ann Arbor, Michigan 48109, USA
[5]Present address: Department of Physics and Astronomy and National Graphene Institute, University of Manchester, Manchester M13 9PL, UK.

[#]These authors contributed equally: Meizhen Huang, Zefei Wu, Xu Zhang.
*Corresponding author. Email: phwang@ust.hk; zymeng@hku.hk; zefei.wu@manchester.ac.uk

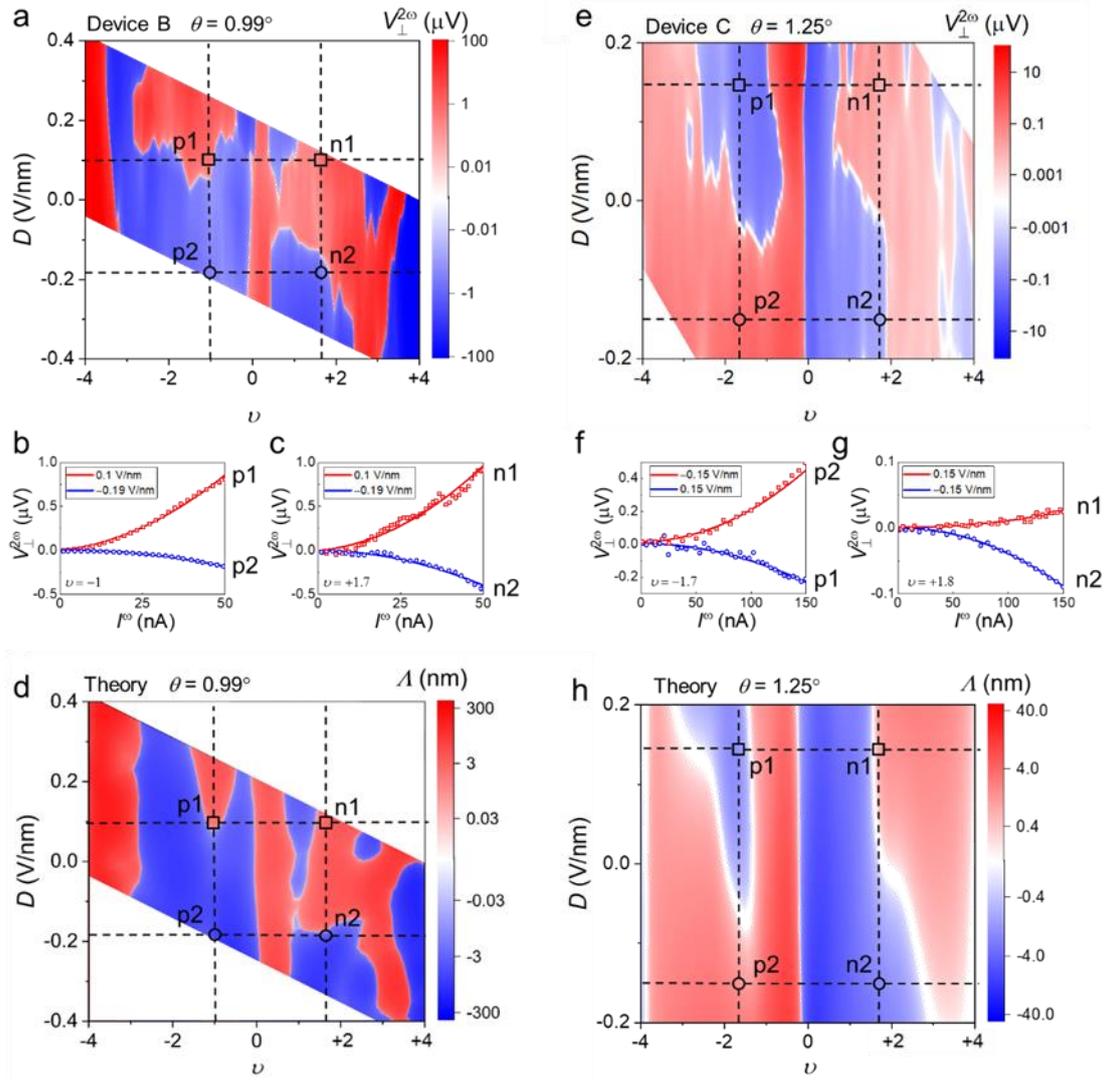

**Supplementary Fig. 1 | Electric field tunable nonlinear Hall effect in different devices.** Data in (**a**)-(**d**) are from device B with a twist angle of 0.99°. **a,** Filling and displacement field mapping of $V_\perp^{2\omega}$ measured at $T = 1.5$ K. **b,** $V_\perp^{2\omega}$ versus $I^\omega$ at various $D$ at $\upsilon = -2.3$. Dots are the experimental data and the solid lines are parabolic fittings. **c,** $V_\perp^{2\omega}$ versus $I^\omega$ at various $D$ at $\upsilon = +1.6$. **d,** Filling and displacement field mapping of $\Lambda$ measured calculated using a strain strength of 0.5% (strain's angle $\theta = 75°$ and dipole's angle $\varphi = 50°$).

Data in (**e**)-(**h**) are from device C with a twist angle of 1.25°. **e,** Filling and displacement field mapping of $V_\perp^{2\omega}$ measured at $T = 1.5$ K. **f,** $V_\perp^{2\omega}$ versus $I^\omega$ at various $D$ at $\upsilon = -1.7$. **g,** $V_\perp^{2\omega}$ versus $I^\omega$ at various $D$ at $\upsilon = +1.0$. **h,** Filling and displacement field mapping of $\Lambda$ measured calculated using a strain strength of 0.3% (strain's angle $\theta = 120°$ and dipole's angle $\varphi = 130°$).

**Supplementary Note 1 | Methods.**

1. <u>Device fabrication</u>: BN and monolayer graphene were exfoliated on SiO$_2$/Si chips. The BN encapsulated TBG was fabricated using a "tear-and-stack" technique using a poly (bisphenol A carbonate) (PC)/polydimethylsiloxane (PDMS) stack on a glass slide[1]. The top gate was first deposited by standard electron-beam lithography and e-beam evaporation. Other rounds of electron-beam lithography and reactive ion etching were applied to create Hall bars and expose graphene edges. Electrical connections were made to the TBG by Cr/Au (3 nm/70 nm) edge contact[2].

2. <u>Electrical characterization</u>: Electronic transport measurements were performed in a cryogenic system which provides stable temperatures ranging from 1.4 to 300 K and fields up to 14 Tesla. Data were obtained with low-frequency lock-in techniques. The current, first and second harmonic signals were measured by Signal Recovery 7280 lock-in amplifier (impedance 100 MΩ).

3. <u>Extraction of carrier density and displacement field</u>: The Hall carrier density as a function of the gate voltage can be extracted using $n_\mathrm{H} = -(1/e)(dR_{xy}/dB)^{-1}$, where $R_{xy}$ is the Hall resistance and $B$ is the perpendicular magnetic field. The gate-induced carrier density $n$ can be determined by a capacitor model: $ne = C_{tg}V_\mathrm{tg} + C_{bg}V_\mathrm{bg} + n_0 e$, where $C_{tg}$ and $C_{bg}$ are the capacitance of the top BN and the bottom SiO$_2$, $n_0$ is carrier density induced by gate offset. The displacement field can be calculated by $D = (C_{bg}V_\mathrm{bg} - C_{tg}V_\mathrm{tg})/2\varepsilon_0$, where $\varepsilon_0$ denotes the vacuum dielectric constant.

**Supplementary Note 2 | Ruling out extrinsic second-harmonic transport effects.**

A second-harmonic transport can arise from both intrinsic (Berry curvature dipole) mechanism and extrinsic (asymmetric contact junction, thermoelectric Joule heating and scattering) mechanism[3-6]. Here we carefully consider those effects and show that the electrically switchable nonlinear Hall effect comes from the intrinsic mechanism.

1. <u>Direction dependence</u>: The intrinsic and extrinsic mechanisms can be distinguished from the magnitude of second harmonic longitudinal responses. The extrinsic mechanisms give rise to a second harmonic longitudinal response with a similar magnitude to the Hall response[5]. In contrast, the longitudinal responses in the experiment are nearly zero at different current injection directions available to the Hall bar geometry (Supplementary Fig. 2). The Hall-dominated signal verifies the Berry curvature dipole induced NHE.

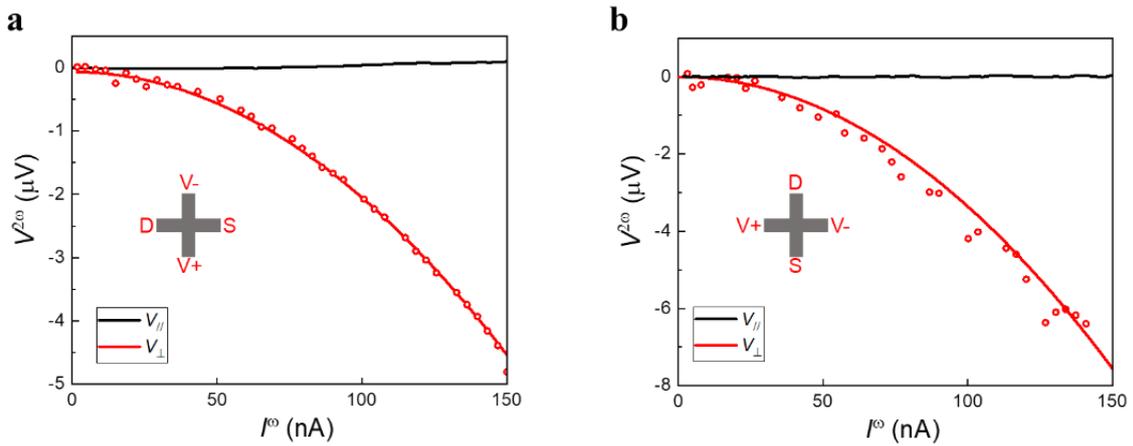

**Supplementary Fig. 2 | Second harmonic responses at different current injection directions. a,** Second harmonic voltages along the longitudinal and transverse directions. Inset: Schematic of the measurement. S represents the source electrode, D represents the drain electrode, and V+, V- represent the Hall voltage measurement electrodes. **b,** Second harmonic responses when current injection and voltage measurement direction are swapped.

Besides, the thermal noise is obtained and shown in Supplementary Fig. 3 by the black dots, it is nearly zero and very small compared to the nonlinear Hall signal.

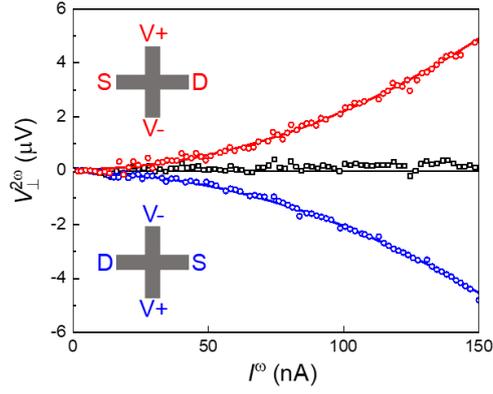

**Supplementary Fig. 3 | Thermal noise obtained marked by black dots from experimental data.** $V_\perp^{2\omega}$ switches sign when the current direction and the voltage probe connection are reversed simultaneously. The thermal noise is obtained by calculating the difference between the red and blue curves.

2. Frequency dependence: Supplementary Fig. 4 shows the frequency dependence of the nonlinear Hall signal. No frequency dependence is observed in the frequency range of 1~1000 Hz. This is consistent with previous experimental observations[4,6] and theoretical calculations[7] of the dipole-induced NHE.

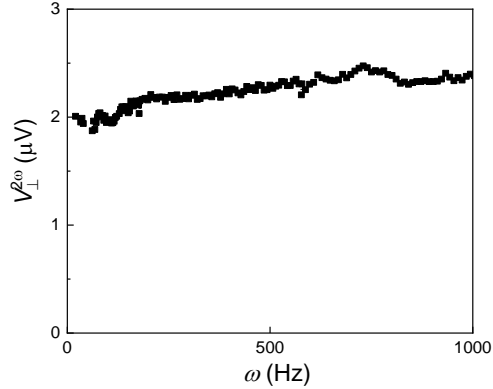

**Supplementary Fig. 4 | The frequency dependence of $V_\perp^{2\omega}$.** Data shown in this figure is measured at $T$ = 1.5 K at $\upsilon = -1$.

3. Filling dependence: Our $V_\perp^{2\omega}$ data show a strong dependence on filling. Especially, $V_\perp^{2\omega}$ show two peaks with opposite signs near the CNP. This feature can be well captured by the theoretical calculation of the Berry curvature dipole based on the Berry curvature hotspots with different signs near the band gap. In contrast, the calculated scattering induced second harmonic transport is sharply different from the experimental observation (calculation details can be found in Supplementary Note 5). As shown in Supplementary Fig. 5c, the scattering induced nonlinear transport shows the same sign for electrons and holes.

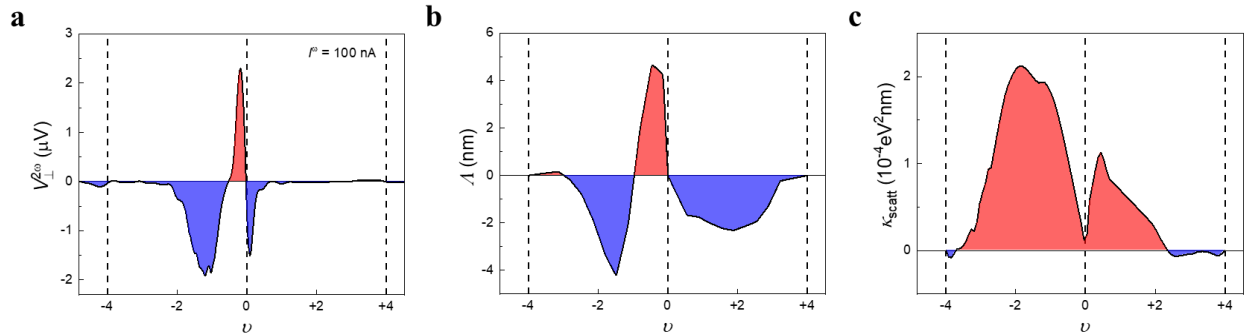

**Supplementary Fig. 5 | Filling dependent nonlinear transport. a,** $V_\perp^{2\omega}$ versus $\upsilon$ measured at $I^\omega = 100$ nA at 1.5K. **b,** Theoretical calculated Berry curvature dipole $\Lambda$ versus $\upsilon$. The observed sign change together with a $V_\perp^{2\omega}$ peak near $\upsilon = -1$ can be fully captured by the theoretical calculation. **c,** Theoretical calculated scattering induced nonlinear signal versus $\upsilon$.

4. <u>Displacement field dependence</u>: The sign change of $V_\perp^{2\omega}$ with the displacement field, which is the highlight of this work, can be captured by the calculated Berry curvature dipole, but cannot be captured by the scattering mechanism. As shown in Supplementary Fig. 6, the contribution of the scattering almost remains a constant at a fixed filling, and never shows any sign change behavior. As a result, the observed electrically switchable nonlinear Hall signal results from the Berry curvature dipole instead of the scattering.

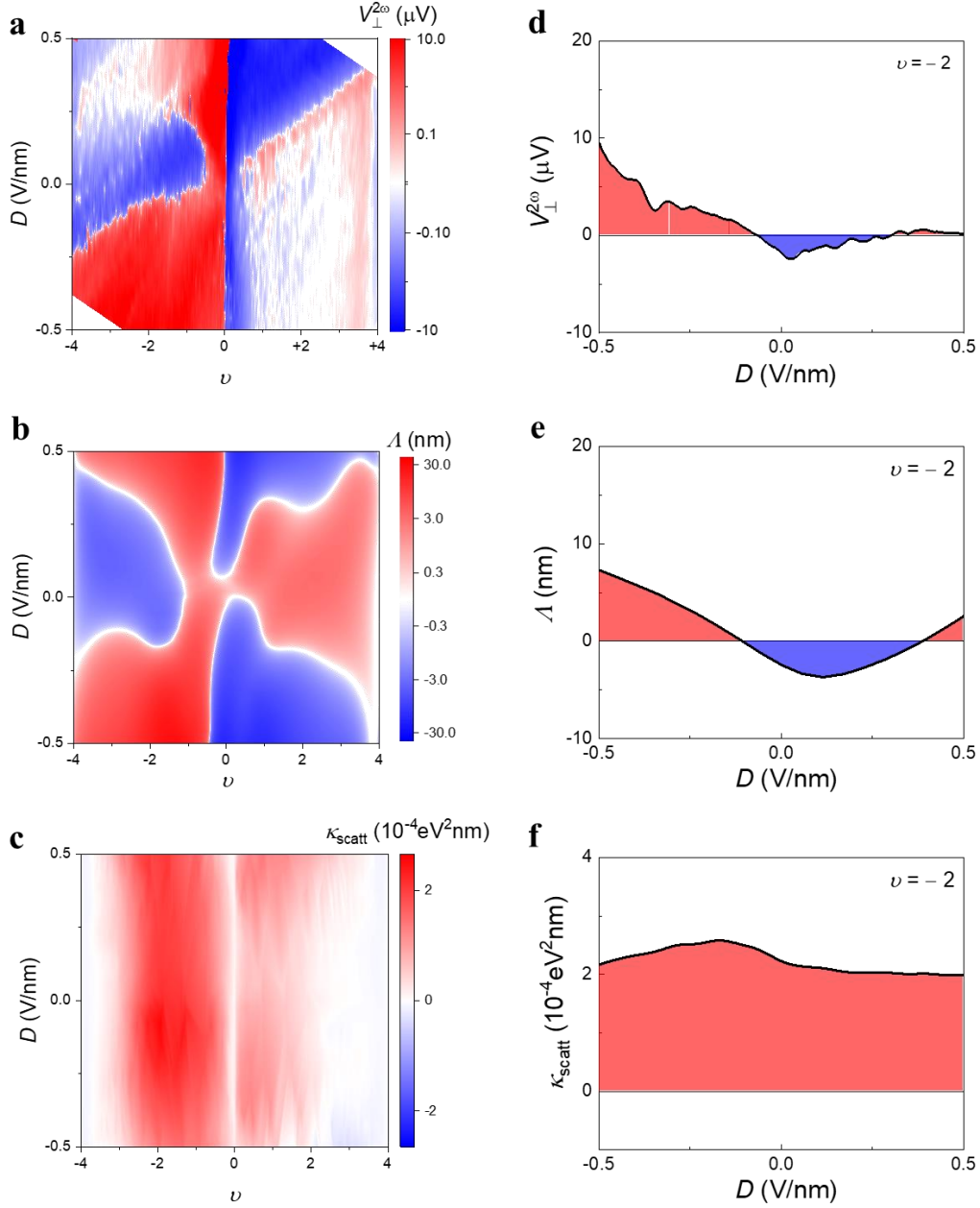

**Supplementary Fig. 6 | Comparison between dipole- and scattering-induced NHE. a-c**, Measured second harmonic Hall voltage (**a**), calculated Berry curvature dipole $\Lambda$ (**b**), and calculated scattering induced nonlinear signal strength (**c**) mapping as functions of the filling $\upsilon$ and the displacement field $D$. **d-f**, Measured second harmonic Hall voltage (**d**), calculated Berry curvature dipole $\Lambda$ (**e**), and calculated scattering strength (**f**) as a function of the displacement field $D$ at filling $\upsilon = -2$.

5. <u>Temperature dependence</u>: The temperature dependence of the nonlinear Hall effect was measured and plotted in Supplementary Fig. 7. $V_\perp^{2\omega}$ depends quadratically on $I^\omega$ at all temperatures and the amplitude of the signal decreases with increasing temperature. To confirm the scaling effect, $V_\perp^{2\omega}/(V_{//}^{\omega})^2$ versus $\sigma_{xx}^2$ is plotted in Supplementary Fig. 7b. Consistent with previous theoretical calculation[3,4], $V_\perp^{2\omega}/(V_{//}^{\omega})^2$ scales linearly with $\sigma_{xx}^2$. Moreover, the Berry curvature dipole calculated from experimental data can be well captured by the theoretical calculation results, but cannot be captured by the scattering contribution (Supplementary Fig. 7c), indicating a robust dipole-induced nonlinear Hall effect.

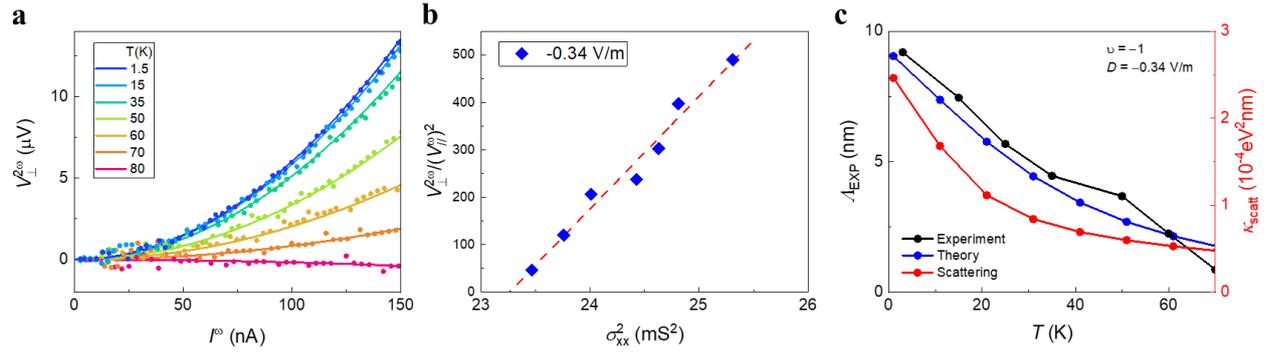

**Supplementary Fig. 7 | Temperature ($T$) dependence of the nonlinear Hall effect at $v = -1$. a**, $V_\perp^{2\omega}$ as a function of $I^\omega$ at different temperatures. Dots are experimental data and lines are parabolic fittings. **b**, $V_\perp^{2\omega}/(V_{//}^{\omega})^2$ as a function of $\sigma_{xx}^2$. Dots are experimental data and the dashed line is a linear fitting. **c**, Berry curvature dipole calculated from experimental data (black), theory (blue), and the scattering strength (red) as a function of temperature.

**Supplementary Note 3 | Displacement field effect on first harmonic transport.**

Previous studies have suggested that the displacement field has little effect on the band structure (or density of states distribution) of TBG [8,9] and that electric field effects in first harmonic transport may arise from disorder[10]. Our experiments on first harmonic transport are consistent with the previous conclusions. As shown in Supplementary Fig. 8a, the charge neutrality gap exists in all experimentally accessible values of displacement field instead of changing the system from insulator to metallic state, which distinguishes with displacement tunable insulating gap in disordered TBG. We also found that the move of the position of Hall density sign-changing point is quite slight when changing the field (balck squares in Supplementary Fig. 8b, c), which is consistent with theoretical calculation (blue line in Supplementary Fig. 8b, c).

We would like to emphasize that even if the first harmonic transport properties of our TBG samples remain relatively unchanged upon the applied displacement field, we find the displacement field indeed can slide the Berry curvature hotspots and lead to significant changes in the nonlinear Hall signal. Our theoretical calculations below will support the idea that the displacement field can induce Berry curvature sliding and switchable Berry curvature dipole in such a strained system.

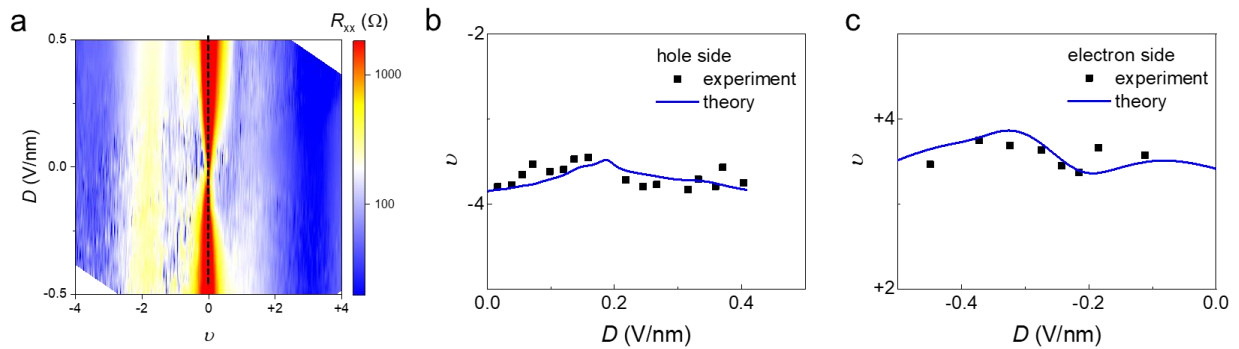

**Supplementary Fig. 8 | Displacement field effect on first harmonic transport. a**, Field and filling dependence of the longitudinal resistance. The charge neutrality gap exists in all experimentally accessible values of displacement field due to strong interlayer hybridization. **b-c**, Field dependence of the Hall density sign-changing point. Experimental results (black squares) can be well described by theoretical calculations (blue line).

**Supplementary Note 4 | Theoretical calculation of Berry curvature dipole.**

Follow references [11-13], we first discuss how to include strain in Bistritzer-MacDonald (BM) model in our calculation. A small amount of strain field is introduced to break the three-fold rotational symmetry. In the theoretical model, we scanned different strain field different orientation and picked the one where the simulated nonlinear Hall data best resembles the experimental results.

Imagine there is a two components vector $x$, then with a small strain $\epsilon$ the new vector $x'$ can be seen as a coordinate transformation matrix $S$ applying on $x$

$$x' = \begin{pmatrix} \frac{1}{1+\epsilon} & 0 \\ 0 & \frac{1}{1-\delta\epsilon} \end{pmatrix} x \approx \begin{pmatrix} 1-\epsilon & 0 \\ 0 & 1+\delta\epsilon \end{pmatrix} x$$

Here $\delta = 0.16$ is the Poisson ratio for graphene. We can rotate the direction of strain with an angle $\theta$ by applying rotation matrixes on $S$

$$S(\epsilon, \theta) = \begin{pmatrix} \cos(\theta) & \sin(\theta) \\ -\sin(\theta) & \cos(\theta) \end{pmatrix} \begin{pmatrix} 1-\epsilon & 0 \\ 0 & 1+\delta\epsilon \end{pmatrix} \begin{pmatrix} \cos(\theta) & -\sin(\theta) \\ \sin(\theta) & \cos(\theta) \end{pmatrix}$$

$$= I + \epsilon \begin{pmatrix} -\cos^2(\theta) + \delta\sin^2(\theta) & (1+\delta)\cos(\theta)\sin(\theta) \\ (1+\delta)\cos(\theta)\sin(\theta) & -\sin^2(\theta) + \delta\cos^2(\theta) \end{pmatrix} \equiv I + \mathrm{E}(\epsilon, \theta)$$

Here $I$ is the $2 \times 2$ unit matrix. Because of the relation $\boldsymbol{a}_M \cdot \boldsymbol{G} = 2\pi$, where $\boldsymbol{a}_M$ is moiré lattice vector and $\boldsymbol{G}$ is moiré reciprocal vector, if the vector $x$ is defined in real space, then the vector in momentum space $p$ should transform as

$$p' = S^{-1}(\epsilon, \theta) p$$

Keep it up to the first order of $\epsilon$

$$S^{-1}(\epsilon, \theta) \approx I - \epsilon \begin{pmatrix} -\cos^2(\theta) + \delta\sin^2(\theta) & (1+\delta)\cos(\theta)\sin(\theta) \\ (1+\delta)\cos(\theta)\sin(\theta) & -\sin^2(\theta) + \delta\cos^2(\theta) \end{pmatrix} \equiv I - \mathrm{E}(\epsilon, \theta)$$

According to Ref. [14], strain will also introduce a fictitious gauge field $\boldsymbol{A}$ for one Dirac cone and $-\boldsymbol{A}$ for the other

$$\boldsymbol{A} = -\frac{\sqrt{3}}{2\alpha a_M} \beta\epsilon(1+\delta) \begin{pmatrix} \cos(2\theta) \\ \sin(2\theta) \end{pmatrix}$$

Here $\beta = 3.14$ is estimated by the first principle calculation for graphene and $\alpha = 1.3°$ is the twisted angle between two graphene layers in our experiment so that moiré lattice constant $a_M \approx \frac{a_0}{\alpha}$.

Without strain, the $K$ points from two twisted layers for valley $\pm$ are labeled by $K_1^\pm, K_2^\pm$ (Supplementary Fig. 9a). We will only focus on valley $+$ below, and valley $-$ can be obtained by time reversal.

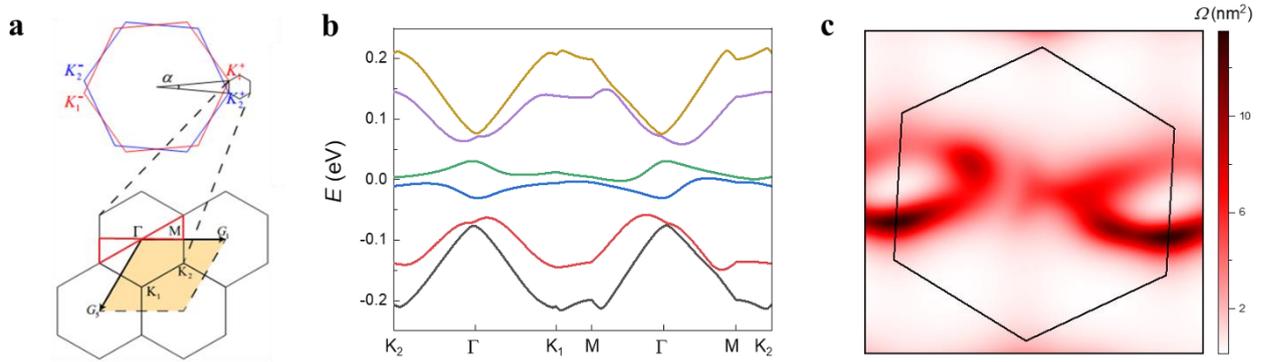

**Supplementary Fig. 9 | Moiré Brillouin zone, band structure and Berry curvature. a,** Schematic moiré Brillouin zone without strain. **b,** Band structure of 6 nearest zero-energy bands along red line in (**a**). **c,** Berry curvature of flat valence band with the black lines correspond to the moiré Brillouin zone.

Considering uniaxial strain $\epsilon_1 = -0.15\%$ for layer 1, $\epsilon_2 = 0.15\%$ for layer 2 and the induced fictitious gauge field $\boldsymbol{A}$, the coordinates for $K_1^+, K_2^+$ are

$$\boldsymbol{K}_1^+ \approx S^{-1}(\epsilon_1, \theta) \frac{G_0}{\sqrt{3}} \begin{pmatrix} 1 \\ \frac{\alpha}{2} \end{pmatrix} + \boldsymbol{A}_1$$

$$\boldsymbol{K}_2^+ \approx S^{-1}(\epsilon_2, \theta) \frac{G_0}{\sqrt{3}} \begin{pmatrix} 1 \\ -\frac{\alpha}{2} \end{pmatrix} + \boldsymbol{A}_2$$

Here $G_0$ is the graphene reciprocal vector length. Keep them up to the first order of $\epsilon$ and $\alpha$ and shift a constant vector $-\frac{G_0}{\sqrt{3}} \begin{pmatrix} 1 \\ 0 \end{pmatrix}$

$$\boldsymbol{K}_1^+ \approx \frac{G_0}{\sqrt{3}} \begin{pmatrix} 0 \\ \frac{\alpha}{2} \end{pmatrix} - \mathrm{E}_1(\epsilon, \theta) \frac{G_0}{\sqrt{3}} \begin{pmatrix} 1 \\ 0 \end{pmatrix} + \boldsymbol{A}_1$$

$$K_2^+ \approx -\frac{G_0}{\sqrt{3}}\begin{pmatrix}0\\ \alpha\\ \frac{1}{2}\end{pmatrix} - E_2(\epsilon,\theta)\frac{G_0}{\sqrt{3}}\begin{pmatrix}1\\0\end{pmatrix} + A_2$$

The second and third terms are contributions from strains. One can see the relative position $K_1^+ - K_2^+$ will not be affected by allocation of total strain $\epsilon = \epsilon_1 - \epsilon_2$, so does any physical result after a momentum shift.

Besides, we may also derive 6 new moiré reciprocal vectors with strain

$$G_i' = S^{-1}(\epsilon,\theta)\begin{pmatrix}1 & -\alpha\\ \alpha & 1\end{pmatrix}G_i - G_i$$

Where $\begin{pmatrix}1 & -\alpha\\ \alpha & 1\end{pmatrix}$ represents the rotation. Keep it up to the first order of $\epsilon$ and $\alpha$, we write all $G_i$ and correspond $G_i'$ explicitly

$$[G_1, G_2, G_3, G_4, G_5, G_6] = G_0\left[\begin{pmatrix}0\\-1\end{pmatrix}, \begin{pmatrix}\frac{\sqrt{3}}{2}\\-\frac{1}{2}\end{pmatrix}, \begin{pmatrix}\frac{\sqrt{3}}{2}\\ \frac{1}{2}\end{pmatrix}, \begin{pmatrix}0\\1\end{pmatrix}, \begin{pmatrix}-\frac{\sqrt{3}}{2}\\ \frac{1}{2}\end{pmatrix}, \begin{pmatrix}-\frac{\sqrt{3}}{2}\\-\frac{1}{2}\end{pmatrix}\right]$$

$$[G_1', G_2', G_3', G_4', G_5', G_6']$$
$$= G_0\alpha\left[\begin{pmatrix}1\\0\end{pmatrix}, \begin{pmatrix}\frac{1}{2}\\ \frac{\sqrt{3}}{2}\end{pmatrix}, \begin{pmatrix}-\frac{1}{2}\\ \frac{\sqrt{3}}{2}\end{pmatrix}, \begin{pmatrix}-1\\0\end{pmatrix}, \begin{pmatrix}-\frac{1}{2}\\-\frac{\sqrt{3}}{2}\end{pmatrix}, \begin{pmatrix}\frac{1}{2}\\-\frac{\sqrt{3}}{2}\end{pmatrix}\right]$$
$$- G_0 E(\epsilon,\theta)\left[\begin{pmatrix}0\\-1\end{pmatrix}, \begin{pmatrix}\frac{\sqrt{3}}{2}\\-\frac{1}{2}\end{pmatrix}, \begin{pmatrix}\frac{\sqrt{3}}{2}\\ \frac{1}{2}\end{pmatrix}, \begin{pmatrix}0\\1\end{pmatrix}, \begin{pmatrix}-\frac{\sqrt{3}}{2}\\ \frac{1}{2}\end{pmatrix}, \begin{pmatrix}-\frac{\sqrt{3}}{2}\\-\frac{1}{2}\end{pmatrix}\right]$$

The second term in $G_i'$ comes from contribution of relative strain.

The BM model without strain is written below[9].

$$H_{k,k'}^+ = \delta_{k,k'}\begin{pmatrix}-\hbar v_F(k - K_1^+)\cdot\sigma & U_0\\ U_0^\dagger & -\hbar v_F(k - K_2^+)\cdot\sigma\end{pmatrix} + \begin{pmatrix}0 & U_1\delta_{k,k'+G_5}\\ U_1^\dagger\delta_{k,k'+G_2} & 0\end{pmatrix}$$
$$+ \begin{pmatrix}0 & U_2\delta_{k,k'+G_6}\\ U_2^\dagger\delta_{k,k'+G_3} & 0\end{pmatrix}$$

Here $2\times 2$ matrix labels two layers, $[U_0, U_1, U_2]$ are also $2\times 2$ matrixes label interlayer hopping.

$$[U_0, U_1, U_2] = \left[\begin{pmatrix} u_0 & u_1 \\ u_1 & u_0 \end{pmatrix}, \begin{pmatrix} u_0 & u_1 e^{-i\frac{2\pi}{3}} \\ u_1 e^{i\frac{2\pi}{3}} & u_0 \end{pmatrix}, \begin{pmatrix} u_0 & u_1 e^{i\frac{2\pi}{3}} \\ u_1 e^{-i\frac{2\pi}{3}} & u_0 \end{pmatrix}\right]$$

Here $2 \times 2$ matrix labels graphene sublattices. $u_0$ and $u_1$ are different hopping amplitudes at AA and AB stacking center. Due to corrugation[15], we take $u_0 = 0.08eV$ and $u_1 = 0.11eV$ from first principles calculation.

Now consider the BM model with strain. One just need to replace all momentum with the vector after the transformation $S^{-1}(\epsilon, \theta)$ and keep the scalar $\boldsymbol{k} \cdot \boldsymbol{\sigma}$ unchanged.

$$H^+_{\boldsymbol{k},\boldsymbol{k}'} = \delta_{\boldsymbol{k},\boldsymbol{k}'} \begin{pmatrix} -\hbar v_F S(\epsilon, \theta)(\boldsymbol{k} - \boldsymbol{K}_1^+) \cdot \boldsymbol{\sigma} & U_0 \\ U_0^\dagger & -\hbar v_F S(\epsilon, \theta)(\boldsymbol{k} - \boldsymbol{K}_2^+) \cdot \boldsymbol{\sigma} \end{pmatrix}$$
$$+ \begin{pmatrix} 0 & U_1 \delta_{\boldsymbol{k},\boldsymbol{k}'+\boldsymbol{G}_5'} \\ U_1^\dagger \delta_{\boldsymbol{k},\boldsymbol{k}'+\boldsymbol{G}_2'} & 0 \end{pmatrix} + \begin{pmatrix} 0 & U_2 \delta_{\boldsymbol{k},\boldsymbol{k}'+\boldsymbol{G}_6'} \\ U_2^\dagger \delta_{\boldsymbol{k},\boldsymbol{k}'+\boldsymbol{G}_3'} & 0 \end{pmatrix}$$

Since $\epsilon$ is small, and the leading order comes from $I$ in $S(\epsilon, \theta) = I + E(\epsilon, \theta)$. BM model we use with small strain can be written as

$$H^+_{\boldsymbol{k},\boldsymbol{k}'} = \delta_{\boldsymbol{k},\boldsymbol{k}'} \begin{pmatrix} -\hbar v_F (\boldsymbol{k} - \boldsymbol{K}_1^+) \cdot \boldsymbol{\sigma} & U_0 \\ U_0^\dagger & -\hbar v_F (\boldsymbol{k} - \boldsymbol{K}_2^+) \cdot \boldsymbol{\sigma} \end{pmatrix} + \begin{pmatrix} 0 & U_1 \delta_{\boldsymbol{k},\boldsymbol{k}'+\boldsymbol{G}_5'} \\ U_1^\dagger \delta_{\boldsymbol{k},\boldsymbol{k}'+\boldsymbol{G}_2'} & 0 \end{pmatrix}$$
$$+ \begin{pmatrix} 0 & U_2 \delta_{\boldsymbol{k},\boldsymbol{k}'+\boldsymbol{G}_6'} \\ U_2^\dagger \delta_{\boldsymbol{k},\boldsymbol{k}'+\boldsymbol{G}_3'} & 0 \end{pmatrix}$$

Here we take $\frac{\hbar v_F}{a_0} = 2.377eV$ and lattice constant for monolayer graphene $a_0 = 0.246nm$ so that $a_M \approx \frac{a_0}{\alpha} = 10.84nm$. In the computation, we take $hBN$ potential[16] $\Delta = 0.017eV$ which is a $\Delta \sigma_z$ between AB sublattices for one layer to break inversion symmetry and open a gap between flat bands.

The contribution from $hBN$ potential in Hamiltonian as

$$H^+_\Delta = \delta_{\boldsymbol{k},\boldsymbol{k}'} \begin{pmatrix} 0 & 0 \\ 0 & \Delta \sigma_z \end{pmatrix}$$

We introduce electric field by considering an electrostatic energy difference $V$ between two layers

$$H_E^+ = \delta_{k,k'} \begin{pmatrix} \frac{V}{2}\sigma_0 & 0 \\ 0 & -\frac{V}{2}\sigma_0 \end{pmatrix}$$

Here $\sigma_0$ is unit matrix and $V$ changes between $\pm 0.4 eV$. The displacement field $D$ is related to the applied electric field, $D = \varepsilon E$, through the out-of-plane dielectric constant $\varepsilon$ of the twisted graphene. Here we suppose weak screen environment between graphene bilayer, therefore $\varepsilon = 1$ is used. The interlayer potential difference $V$ is then determined by $V = Ed$, where $d = 0.35$nm is the interlayer separation. Now we are ready to diagonalize the total Hamiltonian $H^+ = H_{k,k'}^+ + H_\Delta^+ + H_E^+$. The band structure without electric field for 6 nearest zero-energy bands along red line in Supplementary Fig. 9a is shown in Supplementary Fig. 9b.

For computing Berry curvature $\Omega$, we take $100 \times 100$ momentum points in the parallelogram first moiré Brillouin zone as the orange region in Supplementary Fig. 9a. and use the numerical method proposed in [17]. The Berry curvature $\Omega$ without electric field for flat valence band is shown in Supplementary Fig. 9c. Band structure and Berry curvature near charge neutrality with various displacement field is shown in Fig. 3b.

For computing contribution for nonlinear hall conductance near charge neutrality from berry curvature dipole[7], we only consider 6 nearest zero-energy bands, because only bands near fermi surface contribute according to $\partial_b f_0$ in

$$\Lambda_b = \sum_{n=1}^{6} \int_k (\partial_b f_0)\Omega_n$$

In our computation, we tune different strain's angle $\theta$ (the angle between the strain axis and the zigzag direction of graphene) from $0°\sim 180°$ with interval $15°$ and measure dipole $\Lambda_1, \Lambda_5$ along direction $\boldsymbol{G}_1'$ and $\boldsymbol{G}_5'$. We also derived $\Lambda(\varphi)$ along the angular directions $\varphi$ (the angle between the measurement direction and the zigzag direction of graphene) by

$$\partial_\varphi = \cos(\varphi)\partial_x + \sin(\varphi)\partial_y$$
$$\partial_1 = \partial_x$$
$$\partial_5 = \cos(\alpha)\partial_x + \sin(\alpha)\partial_y$$

So that $\partial_\varphi = \cos(\varphi) \partial_1 + \sin(\varphi) \frac{\partial_5 - \cos(\alpha)\partial_1}{\sin(\alpha)}$ and $\Lambda_\varphi = \cos(\varphi) \Lambda_1 + \sin(\varphi) \frac{\Lambda_5 - \cos(\alpha)\Lambda_1}{\sin(\alpha)}$. Here $\alpha$ is the angle between $\boldsymbol{G}'_5$ and $\boldsymbol{G}'_1$ which is approach to $-120°$ with strain. The result in Fig. 3a is computed with strain's angle $\theta = 45°$ and dipole's angle $\varphi = 10°$.

We also introduce how to compute the Hall number for a given tight binding model by semi-classical method below[18]. Consider a nonequilibrium distribution $f(r, k, t)$, then according to relaxation time approximation, the equilibrium particle from last collision contributes to the distribution by $-\frac{\partial f^0(r',k',t')}{\partial \epsilon} \frac{\partial \epsilon}{\partial k} \cdot \frac{dk}{dt'} e^{-(t-t')/\tau} dt'$. In static electromagnetic field, $v = \frac{\partial \epsilon}{\partial k}$, $\frac{dk}{dt'} = -q(v \times B + E)$. Then the nonequilibrium current will be

$$j = -q \int \frac{dk^2}{(2\pi)^2} v(k(0)) f(k(0))$$

$$= q^2 \int \frac{dk^2}{(2\pi)^2} v(k(0)) \left(-\frac{\partial f^0}{\partial \epsilon}\right) \int_{-\infty}^{0} v(k(t)) \cdot (v(k(t)) \times B + E) e^{\frac{t}{\tau}} dt$$

$$= \frac{q^2}{(2\pi)^2} \int \frac{ds}{|v(k(0))|} v(k(0)) \int_{-\infty}^{0} v(k(t)) \cdot E e^{\frac{t}{\tau}} dt$$

In the last line, $ds$ means the integral is along the Fermi surface. According to $j = \sigma E$, we have

$$\sigma_{\alpha\beta} = \frac{q^2}{(2\pi)^2} \int \frac{ds}{|v(k(0))|} v_\alpha(k(0)) \int_{-\infty}^{0} v_\beta(k(t)) e^{\frac{t}{\tau}} dt$$

To compute this integral numerically, we need to find out $\bar{v}_\beta \equiv \int_{-\infty}^{0} v_\beta(k(t)) e^{\frac{t}{\tau}} dt$. The knowledge about $v(k(t))$ is limited that we know $v(k(t)) = \frac{\partial \epsilon}{\partial k}(t)$, but we do not know $\frac{\partial \epsilon}{\partial k}(t)$ at $t$. We need to derive general $v(k(t))$ by using equation of motion $B \times \frac{dk}{dt} = -q(vB^2 + B \times E)$.

$$v(k(t)) = \frac{B \times \left(\frac{dk}{dt} + qE\right)}{-qB^2} = \frac{\partial \epsilon}{\partial k}$$

We can solve the ode for $k(t)$ by knowing $k(0)$. Then replace all the $v(k(t))$ with $k(t)$ in the integral and take small electric field limit

$$\int_{-\infty}^{0} v_\beta(k(t))e^{\frac{t}{\tau}}dt \approx -\frac{1}{qB^2}B \times \int_{-\infty}^{0}\left(\frac{dk}{dt}\right)e^{\frac{t}{\tau}}dt = -\frac{1}{qB^2}B \times \left(k(0) - \int_{-\infty}^{0}\frac{k}{\tau}e^{\frac{t}{\tau}}dt\right)$$

Then the final problem is how to solve the integral $\int_{-\infty}^{0}\frac{k}{\tau}e^{\frac{t}{\tau}}dt$. One should be careful about this part. Since if one takes $\tau \to 0$ limit, the expression $\int_{-\infty}^{0} v_\beta(k(t))e^{\frac{t}{\tau}}dt \to \tau v_\beta(k(0))$ and $\sigma_{\alpha\beta}$ will be identical with the one without magnetic field, where there should be no $\sigma_{xy}$ according to Onsager reciprocity relation. The strong magnetic field limit corresponds to $\tau \to \infty$ limit, where one expects to observe $k(t)$ oscillations in k-space within closed Fermi surface region and evolutions along one direction within open Fermi surface region.

In practice, we compute $k(t)$ numerically. Since there is no explicit expression for energy dispersion in our model, we diagonalize the Hamiltonian to get $\epsilon(k(t) \pm dk)$ at each momentum point and derive $\frac{\partial \epsilon}{\partial k}$. Plug it in $\frac{B \times \left(\frac{dk}{dt}+qE\right)}{-qB^2} = \frac{\partial \epsilon}{\partial k}$ and solve this ode with boundary condition at $k(0) = \frac{\partial \epsilon}{\partial k}$, we can derive the solution for $k(t)$ from a cutoff time $t_c$ to 0. For each point on the Fermi surface, we need to solve this ode once, compute the integral $\int_{t_c}^{0}\frac{k}{\tau}e^{\frac{t}{\tau}}dt$ and then sum over contribution from all the points. In this process, magnetic field $B$ controls the evolution velocity of $k(t)$ in k-space, one need large $B$, a relatively large cutoff time $t_c$ and a small $\tau$ compared with $t_c$ to make sure the cut off integral $\int_{t_c}^{0}\frac{k}{\tau}e^{\frac{t}{\tau}}dt$ converged to the same value as $\int_{-\infty}^{0}\frac{k}{\tau}e^{\frac{t}{\tau}}dt$. Replacing $\alpha, \beta$ with $x$ or $y$, one can derive the conductance and use the relation $\rho = \sigma^{-1}$ to compute $\rho_{xy} = \frac{-\sigma_{xy}}{\sigma_{xx}\sigma_{yy}-\sigma_{xy}\sigma_{yx}}$ and Hall number $n_H = qB\frac{\sigma_{xx}\sigma_{yy}-\sigma_{xy}\sigma_{yx}}{-\sigma_{xy}}$.

**Supplementary Note 5 | Theoretical calculation of scattering induced nonlinear transport.**

In nonmagnetic crystals with inversion-breaking, mobile electrons can exhibit momentum-dependent chirality, which is opposite at k and −k. This chirality leads to skew scattering, which occurs in a similar way as the classical Magnus effect, where a spinning object is deflected when moving in a viscous medium.

Following the similar steps[5,19], we consider skew scattering by taking single gate screened Coulomb potential with charge $e$ impurity at $R_i$ as

$$V_i(|r - R_i|) = \frac{e^2}{4\pi\varepsilon_0\varepsilon}\left(\frac{1}{|r - R_i|} - \frac{1}{\sqrt{|r - R_i|^2 + d^2}}\right)$$

Here $\frac{d}{2}$ is the distance between TBG and gate. In plane wave basis, we have the scattering Hamiltonian $H_i$

$$H_i = \frac{1}{\Omega}\sum_{k,k'}\int dr\, e^{i(k-k')\cdot r}V_i(|r - R_i|)\, c_{k'}^\dagger c_k = \frac{1}{\Omega}\sum_{k,q}e^{iq\cdot R_i}V_i(q)c_k^\dagger c_{k+q}$$

Here $e^{iq\cdot R_i}V_i(q) = e^{iq\cdot R_i}\int dr\, e^{iq\cdot(r-R_i)}V_i(|r - R_i|) = e^{iq\cdot R_i}\frac{e^2}{2\varepsilon_0\varepsilon}\frac{1}{q}(1 - e^{-qd})$ and $\Omega$ is the total area. We would like to rewrite this term in band basis to finish computation of scattering matrix

$$H_i = \frac{1}{\Omega}\sum_{k,q\in mBZ; K,Q} e^{i(q+Q)\cdot R_i}V_i(q + Q)c_{k+K}^\dagger c_{k+q+K+Q}$$

$$= \frac{1}{\Omega}\sum_{k,q\in mBZ; K,Q} e^{i(q+Q)\cdot R_i}V_i(q + Q)\sum_m u_m^*(k + K)c_{k,m}^\dagger \sum_n u_n(k + q + K + Q)c_{k+q,n}$$

$$= \frac{1}{\Omega}\sum_{k,q\in mBZ; Q} e^{i(q+Q)\cdot R_i}V_i(q + Q)\sum_{m,n,K} u_m^*(k + K)u_n(k + q + K + Q)\, c_{k,m}^\dagger c_{k+q,n}$$

$$= \frac{1}{\Omega}\sum_{k,q\in mBZ; Q} e^{i(q+Q)\cdot R_i}V_i(q + Q)\sum_{m,n} \lambda_{m,n}(k, k + q + Q)\, c_{k,m}^\dagger c_{k+q,n}$$

For the first line, we rewrite the summation to whole momentum space by summation in moiré Brillouin zone and summation of all reciprocal vectors. Here we use unitary transformation $c_{k+K}^\dagger = \sum_m u_m^*(k + K)c_{k,m}^\dagger$ and define the form factor $\lambda_{m,n}(k, k + q + Q) = \sum_K u_m^*(k + K)u_n(k + q + K + Q)$. Only considering flat band part and sum over all scattering center $R_i$, one can extract the scattering potential matrix $V_{k,m;k+q,n}$

$$V_{k,m;k+q,n} = \frac{1}{\Omega}\sum_{R_i,Q} e^{i(q+Q)\cdot R_i}V_i(q + Q)\lambda_{m,n}(k, k + q + Q)$$

According to Lippmann-Schwinger formula and Born approximation, we keep the scattering matrix $T$ up to the second order of $V$

$$T = V + V \frac{1}{\epsilon_{k,m} - H_0 + i\mu} V$$

As definition, scattering rate $w_{k,m;k+q,n}$ can be expressed by scattering matrix $T$

$$w_{k,m;k+q,n} = 2\pi |T_{k,m;k+q,n}|^2 \delta(\epsilon_{k,m} - \epsilon_{k+q,n})$$

One can rewrite $w_{k,m;k+q,n} = w^{(S)}_{k,m;k+q,n} + w^{(A)}_{k,m;k+q,n}$ where $w^{(S)}_{k,m;k+q,n}$ is exchanging $k,m$ and $k+q,n$ symmetric and $w^{(A)}_{k,m;k+q,n}$ is antisymmetric. For symmetric contribution of scattering rate, we have Fermi's golden rule with leading order $O(V^2)$

$$w^{(S)}_{k,m;k+q,n} = 2\pi \langle |V_{k,m;k+q,n}|^2 \rangle_{dis} \delta(\epsilon_{k,m} - \epsilon_{k+q,n})$$

But this symmetric scattering will not contribute to skew scattering, one need to compute antisymmetric scattering rate $w^{(A)}_{k,m;k+q,n}$ with leading order $O(V^3)$

$$w^{(A)}_{k,m;k+q,n} = i2\pi^2 \sum_{p,l} \langle V_{k,m;k+q,n} V_{k+p,l;k,m} V_{k+q,n;k+p,l} - c.c. \rangle_{dis} \delta(\epsilon_{k,m} - \epsilon_{k+p,l}) \delta(\epsilon_{k,m} - \epsilon_{k+q,n})$$

$$= -(2\pi)^2 \sum_{p,l} Im(\langle V_{k,m;k+q,n} V_{k+p,l;k,m} V_{k+q,n;k+p,l} \rangle_{dis}) \delta(\epsilon_{k,m} - \epsilon_{k+p,l}) \delta(\epsilon_{k,m} - \epsilon_{k+q,n})$$

$$= -(2\pi)^2 \sum_{p,l \in (\epsilon_{k,m} \pm \Delta)} \frac{Im(\langle V_{k,m;k+q,n} V_{k+q,n;k+p,l} V_{k+p,l;k,m} \rangle_{dis})}{2\Delta} \delta(\epsilon_{k,m} - \epsilon_{k+q,n})$$

Here $\langle \ \rangle_{dis}$ is the distribution average for scattering center $R_i$, $\Delta$ is small and the summation only for states which within the energy region $\epsilon_{k,m} \pm \Delta$ and we have used the equation below to derive the formula above

$$\lim_{\mu \to 0} Im\left(\frac{1}{\epsilon - H_0 + i\mu}\right) = i\pi \delta(\epsilon - H_0)$$

We assume different scattering centers are random distribution, then the summation

$$\langle \sum_{R_{1,i}, R_{2,i}, R_{3,i}} e^{i(q_1+Q_1) \cdot R_{1,i}} e^{i(q_2+Q_2) \cdot R_{2,i}} e^{i(q_3+Q_3) \cdot R_{3,i}} \rangle_{dis} \approx N_i \delta_{q_1+Q_1+q_2+Q_2+q_3+Q_3, Q}$$

Here we only keep the leading order of impurity density $\frac{N_i}{\Omega}$ then the contribution is just three $V$ matrixes production keeping momentum and energy conservation. Ignore side jump, and finally we can put this $w^{(A)}_{k,m;k+q,n}$ in the skew scattering conductance coming from Boltzmann equation

$$\partial_t f_l + qE_\alpha e^{iwt} \partial_\alpha f_l = -\sum_{l'} (w_{l'l} f_l - w_{ll'} f_{l'})$$

Here $l$ is the label for state with momentum $k$ and band $m$. We expand $w_{ll'}$ and $f_l$ up to leading order by seeing antisymmetric scattering rate $w^{(A)}_{k,m;k+q,n}$ and electric field $\boldsymbol{E} = E_\alpha e^{iwt}$ as perturbation

$$w_{ll'} = w^{(S)}_{ll'} + w^{(A)}_{ll'}$$

$$f_l = f_l^{(0,0)} + f_l^{(1,0)} + f_l^{(1,1)} + f_l^{(2,0)} + f_l^{(2,1)}$$

Here $f_l^{(n_1,n_2)}$ means distribution function of order $E^{n_1}$ and $(w^{(A)})^{n_2}$ and $f_l^{(0,0)} = \frac{1}{e^{\beta(\epsilon-\mu)}+1}$ is fermi distribution. According to optical theorem, we have

$$\sum_{l'} w_{ll'} = 2\pi \sum_{l'} |T_{l;l'}|^2 \delta(\epsilon_l - \epsilon_{l'}) = 2\pi \sum_{l'} |T_{l';l}|^2 \delta(\epsilon_l - \epsilon_{l'}) = \sum_{l'} w_{l'l}$$

This means $\sum_{l'} w^{(A)}_{ll'} = 0$. By writing Boltzmann equation order by order, we have recursive relation

$$\partial_t f_l^{(1,0)} + qE_\alpha e^{iwt} \partial_\alpha f_l^{(0,0)} = -\sum_{l'} w^{(S)}_{l'l} \left(f_l^{(1,0)} - f_{l'}^{(1,0)}\right)$$

$$\partial_t f_l^{(1,1)} = -\sum_{l'} w^{(S)}_{l'l} \left(f_l^{(1,1)} - f_{l'}^{(1,1)}\right) + \sum_{l'} w^{(A)}_{ll'} f_{l'}^{(1,0)}$$

$$\partial_t f_l^{(2,0)} + qE_\alpha e^{iwt} \partial_\alpha f_l^{(1,0)} = -\sum_{l'} w^{(S)}_{l'l} \left(f_l^{(2,0)} - f_{l'}^{(2,0)}\right)$$

$$\partial_t f_l^{(2,1)} + qE_\alpha e^{iwt} \partial_\alpha f_l^{(1,1)} = -\sum_{l'} w^{(S)}_{l'l} \left(f_l^{(2,1)} - f_{l'}^{(2,1)}\right) + \sum_{l'} w^{(A)}_{ll'} f_{l'}^{(2,0)}$$

Take relaxation time approximation

$$\sum_{l'} w^{(S)}_{l'l} \left(f_l^{(n_1,n_2)} - f_{l'}^{(n_1,n_2)}\right) = \frac{f_l^{(n_1,n_2)}}{\tau^{(n_1,n_2)}}$$

We can solve recursive equations self-consistently

$$f_l^{(1,0)} = \frac{-q\tau^{(1,0)}}{1+iw\tau^{(1,0)}} E_\alpha e^{iwt} \partial_\alpha f_l^{(0,0)}$$

$$f_l^{(1,1)} = \frac{\tau^{(1,1)}}{1+iw\tau^{(1,1)}} \sum_{l'} w_{ll'}^{(A)} f_{l'}^{(1,0)}$$

$$f_l^{(2,0)} = \frac{-q\tau^{(2,0)}}{1+i2w\tau^{(2,0)}} E_\alpha e^{iwt} \partial_\alpha f_l^{(1,0)}$$

$$f_l^{(2,1)} = \frac{\tau^{(2,1)}}{1+i2w\tau^{(2,1)}} \sum_{l'} w_{ll'}^{(A)} f_{l'}^{(2,0)} + \frac{-q\tau^{(2,1)}}{1+i2w\tau^{(2,1)}} E_\alpha e^{iwt} \partial_\alpha f_l^{(1,1)}$$

According to the definition of current operator and velocity operator

$$j_\alpha = q\sum_{l_m} \int d^2k\, v_\alpha f_l = \sigma_{\alpha\beta} E_\beta + \chi_{\alpha\beta\gamma} E_\beta E_\gamma$$

$$v_\alpha = \partial_\alpha \epsilon - q(\mathbf{E}\times\mathbf{\Omega})_\alpha$$

The dominant contribution from skew scattering comes from $f_l^{(1,1)}$ and $f_l^{(2,1)}$

$$\sigma_{\alpha\beta}^{(skew)} = \frac{q\tau^{(1,1)}}{1+iw\tau^{(1,1)}} \frac{-q\tau^{(1,0)}}{1+iw\tau^{(1,0)}} \sum_{l_m}\int d^2k\, \frac{\partial\epsilon_l}{\partial k_\alpha} \sum_{l'} w_{ll'}^{(A)} \partial_\beta' f_{l'}^{(0,0)}$$

$$\chi_{\alpha\beta\gamma}^{(skew,1)} = \frac{q\tau^{(2,1)}}{1+i2w\tau^{(2,1)}} \frac{-q\tau^{(2,0)}}{1+iw\tau^{(2,0)}} \frac{-q\tau^{(1,0)}}{1+iw\tau^{(1,0)}} \sum_{l_m}\int d^2k\, \frac{\partial\epsilon_l}{\partial k_\alpha} \sum_{l'} w_{ll'}^{(A)} \partial_\beta' \partial_\gamma' f_{l'}^{(0,0)}$$

$$\chi_{\alpha\beta\gamma}^{(skew,2)} = \frac{-q^2\tau^{(2,1)}}{1+i2w\tau^{(2,1)}} \frac{\tau^{(1,1)}}{1+iw\tau^{(1,1)}} \frac{-q\tau^{(1,0)}}{1+iw\tau^{(1,0)}} \sum_{l_m}\int d^2k\, \frac{\partial\epsilon_l}{\partial k_\alpha} \partial_\beta \sum_{l'} w_{ll'}^{(A)} \partial_\gamma' f_{l'}^{(0,0)}$$

Here $l_m$ means the summation to band label in $l$. If we assume $w\to 0$ and $\tau^{(n_1,n_2)} \to \tau$, we have

$$\sigma_{\alpha\beta}^{(skew)} = (\pi q\tau)^2 \sum_m \int d^2k\, \frac{\partial\epsilon_m(k)}{\partial k_\alpha} \sum_{p,l;q,n\in(\epsilon_{k,m}\pm\Delta)} \frac{Im(\langle V_{k,m;k+q,n} V_{k+q,n;k+p,l} V_{k+p,l;k,m}\rangle_{dis})}{\Delta^2} \partial_\beta' f_n^{(0,0)}(k+q)$$

$$\chi_{\alpha\beta\gamma}^{(skew,1)} = -\pi^2(q\tau)^3 \sum_m \int d^2k\, \frac{\partial\epsilon_m(k)}{\partial k_\alpha} \sum_{p,l;q,n\in(\epsilon_{k,m}\pm\Delta)} \frac{Im(\langle V_{k,m;k+q,n} V_{k+q,n;k+p,l} V_{k+p,l;k,m}\rangle_{dis})}{\Delta^2} \partial_\beta' \partial_\gamma' f_n^{(0,0)}(k+q)$$

$$\chi_{\alpha\beta\gamma}^{(skew,2)} = -\pi^2(q\tau)^3 \sum_m \int d^2k\, \frac{\partial\epsilon_m(k)}{\partial k_\alpha} \partial_\beta \sum_{p,l;q,n\in(\epsilon_{k,m}\pm\Delta)} \frac{Im(\langle V_{k,m;k+q,n} V_{k+q,n;k+p,l} V_{k+p,l;k,m}\rangle_{dis})}{\Delta^2} \partial_\gamma' f_n^{(0,0)}(k+q)$$

Here $\partial'_\beta$ means the partial derivative at band $n$ and momentum $k+q$. These expressions are the form easy to compute numerically. One can notice only bands near Fermi surface will contribute according to $\partial'_\beta f_n^{(0,0)}(k+q)$. We take $m$ as two flat bands in TBG. By setting an energy window $\Delta$, for a given $k$ we compute $\frac{\partial \epsilon_m(k)}{\partial k_\alpha}$ and collect all states within the energy window. By iterating collected states, we can compute the rest part in the integral. The result is shown in Supplementary Figure 10.

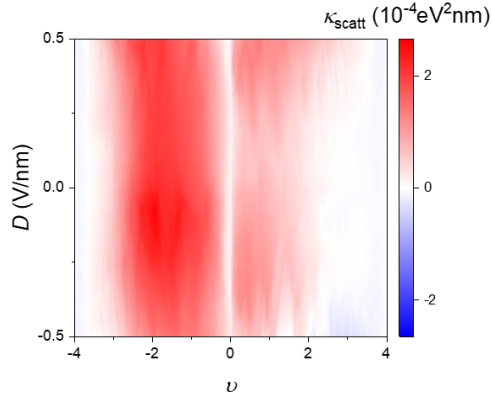

**Supplementary Fig. 10 | Filling and displacement field dependent scattering strength.**

To compare the scattering contribution with the contribution from BCD

$$\chi_{\alpha\beta\gamma}^{(BCD)} = -q^3 \tau \epsilon_{\alpha\beta\eta} \int f^{(0,0)} \partial_\gamma \Omega_\eta$$

We need to compute the coefficient $C = 3.464 \times 10^{-10} \tau^2 \approx 8 \times 10^{-4}$, where $3.464 \times 10^{-10}$ is the size of the total simulated area, and we take $\tau \approx 10^{-12}/6.582 \times 10^{-16}$ eV$^{-1}$ as a constant for convenience.

This means the contribution from skew scattering corresponds to the contribution of $CN_i \kappa_{scatt}$ (nm) BCD. For example, at filling $\upsilon = -1$, displacement field $D = -0.34$ V/nm, $\kappa_{scatt} = 1.6809 \times 10^{-4}$ eV$^2 nm$, if the disorder density is $N_i = 0.1 \times 10^9$ cm$^{-2}$, the effective dipole can be calculated to be 13.4nm.

Since we observe near-zero nonlinear signal along longitudinal direction where Berry curvature dipole cannot contribute but disorder has little directional bias, we can use this observed value to estimate the effective disorder density. From Supplementary Fig 2, disorder-induced second harmonic longitudinal signal in our sample is ~0.04 μV at ~100 nA, which equals the contribution from a 0.37nm BCD. The effective disorder density extracted with our computation is ~2.7× 10$^6$ cm$^{-2}$. In contrast, the second harmonic longitudinal signal is 3000-4000 μV at ~1 μA in a TBG sample with relatively low mobility (Figure. S3 in Ref. [20]). The corresponding nonlinear voltage generation efficiency $V^{2\omega}/(I^{\omega})^2$ is about 100 times larger than in our high-mobility sample, leading to the estimated effective disorder density to be ~$3.4\times 10^8$ cm$^{-2}$.

**Supplementary reference**